# Plasma and light: Shaping High Harmonic Generation in Time-Varying Media

W. Jaffray[1], M. A. Vincenti[2], D. de Ceglia[2], C. Rizza[3],
N. Litchinitser[4], J. Trull[5], and C. Cojocaru[5], M. Ferrera[1], M. Scalora[1]

[1]Institute of Photonics and Quantum Sciences, Heriot-Watt University, SUPA, Edinburgh EH144AS, United Kingdom

[2]Department of Information Engineering, University of Brescia, 25123 Brescia, Italy

[3]Department of Physical and Chemical Sciences, University of L'Aquila, I-67100 L'Aquila, Italy

[4]Department of Electrical and Computer Engineering, Duke University, Durham, North Carolina 27708, USA

[5]Department of Physics, Universitat Politècnica de Catalunya, 08222 Terrassa (Barcelona), Spain

## Abstract

High harmonic generation in solids is typically described under the assumption of a stationary medium. This assumption is rarely questioned because the electronic properties of most media evolve on timescales much slower than the harmonic emission process itself. However, in strongly driven time-varying transparent conductive oxides the plasma frequency, carrier density, and effective mass can evolve on the same femtosecond timescale as harmonic generation. Under these conditions, harmonic generation and material evolution are no longer independent processes. Within this context, we consider a two-field pumping scheme in which a strong field generates a comb of high-order harmonics across a broad spectral window, while a weaker field simultaneously reshapes its own individual spectral lines. Distinct from geometry-dependent wave-mixing processes, the evolving electron plasma reshapes multiple collinear harmonic channels simultaneously. A fully self-consistent hydrodynamic-Maxwell framework reveals that even- and odd-order harmonics naturally emerge from free- and bound-electron dynamics without the typical need to invoke phenomenological nonlinearities. Our model captures the full spectral evolution on an ultra-broad bandwidth, while experiments performed on ITO confirm the distinct behavior of bulk-driven odd harmonics and interface-driven even harmonics under strong pumping. These results establish time-dependent plasma dynamics as a powerful new tool for broadband hyperspectral shaping of multiple harmonic channels beyond the framework of four-wave-mixing processes. Our findings are of particular relevance to ultrafast science, with potential applications in ultrafast pulse manipulation and the generation and control of tailored multi-harmonic waveforms.

## Introduction

High harmonic generation (HHG) in solids has been observed in bulk crystals, semiconductors, atomically thin materials, plasmonic structures, and dielectric metasurfaces [1-8].

Time-varying photonic systems driven by multiple beams can unlock unusual harmonic responses [9]. When an optical pump drives a centrosymmetric epsilon-near-zero (ENZ) material [10-18] out of equilibrium [19], the electron fluid enters an unusual dynamical regime, and a control wave experiences a time-varying response where harmonic signals can emerge with enhanced efficiency.

It is important to note that standard multi-beam interactions in ultrathin nonlinear media are conventionally exploited through wave-mixing processes, where the frequencies and propagation directions of the generated signals are determined by energy and momentum conservation [20-21]. Here we explore a fundamentally different regime, in which an intense optical field (the pump) generates a high-harmonic comb [22-23], while a second, weaker field (the control wave) reshapes its spectral lines through a pump-induced time-dependent material response. The effect, which is studied within subwavelength transparent conducting films, is therefore not defined by a specific wave-mixing geometry [24-25], but by the ultrafast evolution of the material electron plasma. Importantly, the response of each harmonic channel can be independently sensitive to the intensity, polarization and temporal properties of the interacting fields, providing multiple degrees of freedom for simultaneously tailoring the spectral structure of the harmonic comb [26]. This establishes a novel route towards hyperspectral shaping of high-harmonic emission in time-varying media. While recent work has demonstrated control and enhancement of HHG through electronic structure [27-28], dimensionality [4], plasmonic confinement [6], and dielectric metasurfaces [5], the present approach exploits a fundamentally different mechanism in which the material parameters themselves evolve on the same timescale as the harmonic emission process. Recent studies have demonstrated all-optical modulation of HHG in solids through carrier-population control [29], excitation-induced dephasing and phase transitions [30], and dynamically shifting resonances [31]. However, these works largely interpret the observations within semiconductor Bloch theory, carrier-relaxation, or cavity-resonance frameworks and do not address the electrodynamics of a time-evolving electron plasma. In particular, they do not consider ENZ media, surface-Coulomb-driven even harmonics, magnetic and convective nonlinearities, hydrodynamic pressure, or the decomposition of bulk and interfacial harmonic sources. By contrast, the present work examines HHG in a strongly driven ENZ conductor whose plasma frequency evolves on the same femtosecond timescale as the harmonic emission process, causing the medium itself to become an active dynamical participant in frequency conversion.

A nonlinear Lorentz-oscillator model has been employed to represent the interfacial second-order response in time-varying centrosymmetric ENZ materials [32]. This approach can reproduce overall conversion efficiencies under specific experimental conditions, but its phenomenological nature makes it challenging to capture the full hierarchy of microscopic contributions that arise in strongly driven, time-varying media [33-37]. Models that neglect surface Coulomb nonlinearities, magnetic-dipole sources, convective hydrodynamic terms, and nonlocal pressure [38] may miss important physical mechanisms. These considerations motivate the use of a more comprehensive, self-consistent framework capable of describing both free- and bound-electron dynamics under ultrafast excitation.

To overcome these limitations, we use a self-consistent hydrodynamic-Maxwell framework that couples Drude-Lorentz linear and nonlinear dispersions of both free and bound electrons with several key parameters including hydrodynamic pressure, viscosity, Coulomb and magnetic Lorentz forces, intensity-dependent carrier density and effective mass [38-39]. Our model naturally decomposes harmonic sources into bulk, surface, magnetic, convective, and nonlocal contributions and preserves full k- and ω-dependent linear and nonlinear dispersions. This rigorous theoretical treatment allows for a predictive study of high harmonic generation in time-varying systems while addressing the persistent challenge of explaining the origin of even-order harmonic generation in centrosymmetric materials like indium tin oxide (ITO), aluminum zinc oxide (AZO), and several other TCOs. The central question we address is therefore not simply how harmonics are generated in a time-varying medium, but whether the harmonic emission itself becomes a “probe” of the medium's dynamically evolving plasma. If so, spectral broadening, line splitting, frequency shifting, and pulse compression should emerge naturally as signatures of the underlying plasma evolution rather than as isolated nonlinear effects.

**Pump-Induced Modification of the Material Response**

Fig. 1a illustrates the excitation scheme where an ultrafast, TM-polarized pump and a time-delayed TM-polarized control wave impinge non collinearly and co-propagate through a 25 nm-thick ITO film. The pump pulse, with a peak intensity of 1 $TW/cm^2$ and duration of 12 fs (full width at half-maximum), is tuned to a carrier wavelength of 800 nm and impinges normally on the sample generating harmonics of odd order. The control wave pulse, 100 $GW/cm^2$, 43 fs duration, centered at 1200 nm, arrives at a 45° angle relative to the surface normal and is synchronized to overlap with the pump inside the film’s interior.

In Fig.1b we present the local dielectric constant of our ITO layer as a function of wavelength. Here the real part of the permittivity crosses through zero at roughly 1235 nm, marking the ENZ point. For a static plasma frequency and local dielectric response, around this ENZ crossing field confinement intensifies, making the material exceptionally sensitive to even modest changes in plasma frequency.

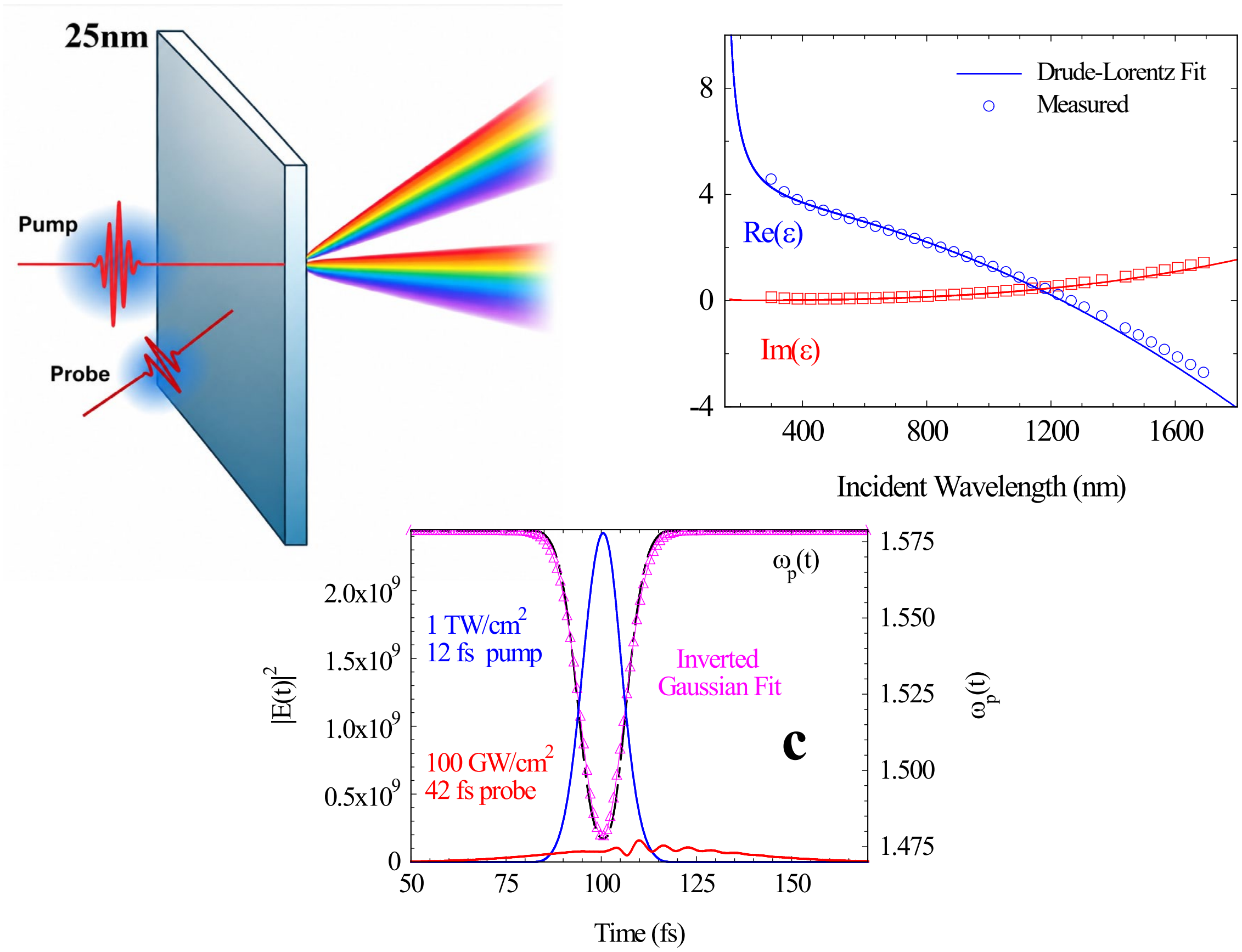

**Fig.1:** (a) Schematic representation of the interaction between a strong pump and weaker control wave with a 25 nm-thick ITO layer. (b) Local, complex dielectric function of the ITO under consideration. Markers represent measured data. The solid lines are the Drude-Lorentz fit we use. (c) Pump (solid blue curve) and control wave (solid red curve) electric field intensities extracted in the middle of the ITO layer. The black curve represents $\omega_p(\mathbf{r},t)$ for the control wave field, as expressed in Eq. S2. The purple dashed curve with markers is a nearly seamless gaussian fit of the plasma frequency.

Fig. 1c details the results of a simulation of the ultrafast pump-control wave dynamics inside the film. Plotted on the left y-axis are the temporal envelopes of the pump (blue curve) and control wave (red curve) electric field intensities, along with the instantaneous plasma frequency of conduction electrons experienced simultaneously by both fields (black curve, right y-axis) near the film's center. The significance of Fig. 1c extends beyond the particular pump-probe configuration considered here. It demonstrates that the material parameters cannot always be regarded as static spectators during harmonic generation. Instead, the plasma itself becomes a dynamical participant whose evolution can reshape the spectrum emitted by the probe. Significantly, our simulations verify that these spatial profiles maintain excellent uniformity across the film thickness, a characteristic that persists even in layer thicknesses exceeding several tens of nanometers. For our excitation conditions, the plasma frequency shifts by ~10% at peak pump intensity, tracking the pump's temporal profile, and closely resembles an inverted Gaussian. We note that it is possible for the plasma frequency to display multiple peaks/valleys depending on parameter choices [39]. A more detailed justification for the temporal shape of the plasma-frequency modulation is provided in the Supplementary Material. These calculations show that dynamics drive the plasma frequency away from a single inverted Gaussian and toward the simple double-dipped envelope used in our harmonic generation simulations to compare with our experimental results.

The modulation of the plasma frequency thus occurs unaffected by harmonic generation processes, which as we will see is instead strongly affected. Therefore, the plasma frequency can be incorporated directly as a time-dependent parameter in the control wave's material equations of motion, alongside the nonlinear polarization terms responsible for harmonic emission to capture the new material response and frequency-conversion dynamics. The ultrafast evolution of the plasma frequency in Fig. 1c provides a clear guide for determining when the modulation will have its strongest impact on the generated harmonics. Our simulations identify several conditions that optimize/maximize the effect: (i) the pump pulse should be significantly more intense, and somewhat shorter, than the control wave; (ii) all else being equal, thinner layers isolate material properties from geometrical resonances, and yield a more pronounced response; and (iii) for typical TCO parameters, the peak power density should be sufficient to induce at least a 10% modulation of the plasma frequency, which corresponds to a pump of roughly 1 $TW/cm^2$, as shown in Fig. 1.

**Control wave -Mediated Material Modification**

We simulate nonlinear HHG using the model described in detail in Refs. [38-39], modified here to account for the time-varying material response discussed above. The free-electron polarization includes surface Coulomb terms, magnetic Lorentz forces, convective nonlinearities, nonlocal pressure and viscosity, and hot-electron corrections, while in addition to accurate linear and nonlinear dispersions, the bound-electron oscillators contribute both surface and intrinsic bulk nonlinearities. The time-dependent electronic response induced by the stronger field is incorporated directly into these material dynamics, allowing the generated harmonics to evolve within the ultrafast-modulated electron plasma. In this way, even- and odd-order harmonics emerge naturally from their underlying microscopic sources rather than through phenomenologically imposed nonlinear susceptibilities. The complete equations of motion and numerical implementation are provided in the Supplementary Material.

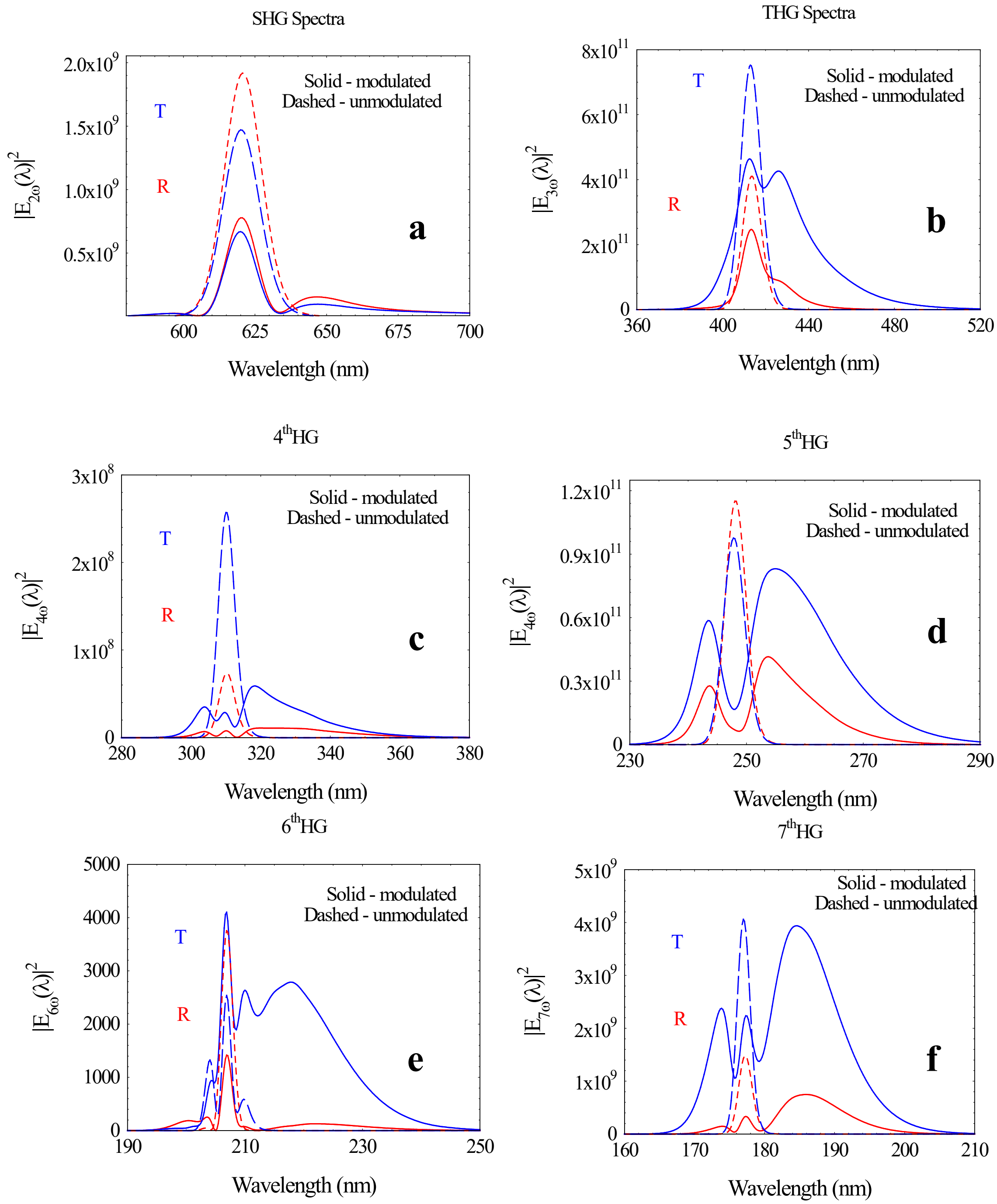

**Fig.2:** A ~43 fs, 190 GW/cm$^2$ pump pulse is incident at 45° on the 25 nm-thick ITO layer shown in Fig.1a. The instantaneous plasma frequency is modulated to mimic a ~2-3 TW/cm$^2$ incident strong pump pulse, resulting in transmitted and reflected spectra for: (a) SHG; (b) THG; (c) 4$^{th}$ HG; (d) 5$^{th}$ HG; (e) 6$^{th}$ HG; and (f) 7$^{th}$ HG. The spectra that result from having a static plasma frequency are relatively narrow-band, except for the transmitted 6$^{th}$ harmonic signal, which shows side bands in transmission.

In Fig. 2 we show the resulting multi-spectral line evolution for the 25 nm ITO layer shown in Fig.1a. The plot compares transmitted and reflected spectra of 2nd through 7th harmonics, under both static plasma frequency and ultrafast time-dependent modulation. The incident control wave pulse has a peak power density of 190 GW/cm² tuned to 1240 nm and impose a free-carrier plasma frequency $\omega_p(t)$ whose inverted Gaussian temporal profile (~12 fs duration) mirrors that in Fig.1c. As the pulse reaches the film midpoint, $\omega_p(t)$ is redshifted by ~20% from its equilibrium value.

The insertion of $\omega_p(t)$ into the control wave's Drude polarization term mimics the action of a 12 fs pump pulse with peak power density of order ~2-3 TW/cm², capturing both the instantaneous refractive-index change seen by the control wave and its harmonics as well as the resulting nonlinear frequency conversion, in a self-consistent crossing of the ITO layer. As pointed out earlier, given the film's subwavelength thickness, we treat $\omega_p(t)$ as spatially uniform across the interaction volume, a condition easily verified using our model.

It is useful at this stage to pause and ask what a harmonic field actually experiences while propagating through such an evolving plasma. In a stationary medium each harmonic is generated under a single phase-matching condition. Here that condition continuously changes during the emission process. Ultrafast modulation of the plasma frequency imprints a rapid, time-dependent phase shift onto the control wave and all its harmonics, that manifests in at least *four* clear spectral and temporal signatures, in different ways for even and odd harmonics. The first predicted effect is *spectral broadening*: since $\omega_p(t)$ follows an inverted Gaussian with ~12 fs width, the refractive index seen by the harmonic fields is continuously sweeping during emission. In the time domain this is equivalent to imposing an ultrafast phase chirp on each harmonic. Just as the Fourier-transform reveals broadening of the spectral envelope of a chirped pulse, (longer wavelengths appear where $\omega_p(t)$ is most red-shifted, shorter wavelengths where it hasn't yet shifted) so every harmonic line fans out into a generally broader band.

The second predicted effect is related to spectral broadening and is *compression of the generated harmonic pulses:* in Fig. 3 we display the predicted, generated temporal profiles of 3$^{rd}$ and 5$^{th}$ harmonic signal intensities at the medium's exit, with and without temporal plasma frequency modulation. Modulating the plasma frequency on 12 fs time scale reduces pulse duration from approximately 25fs in the static case, down to approximately 5 fs in the time dependent case. The 7th harmonic also shows similar behavior.

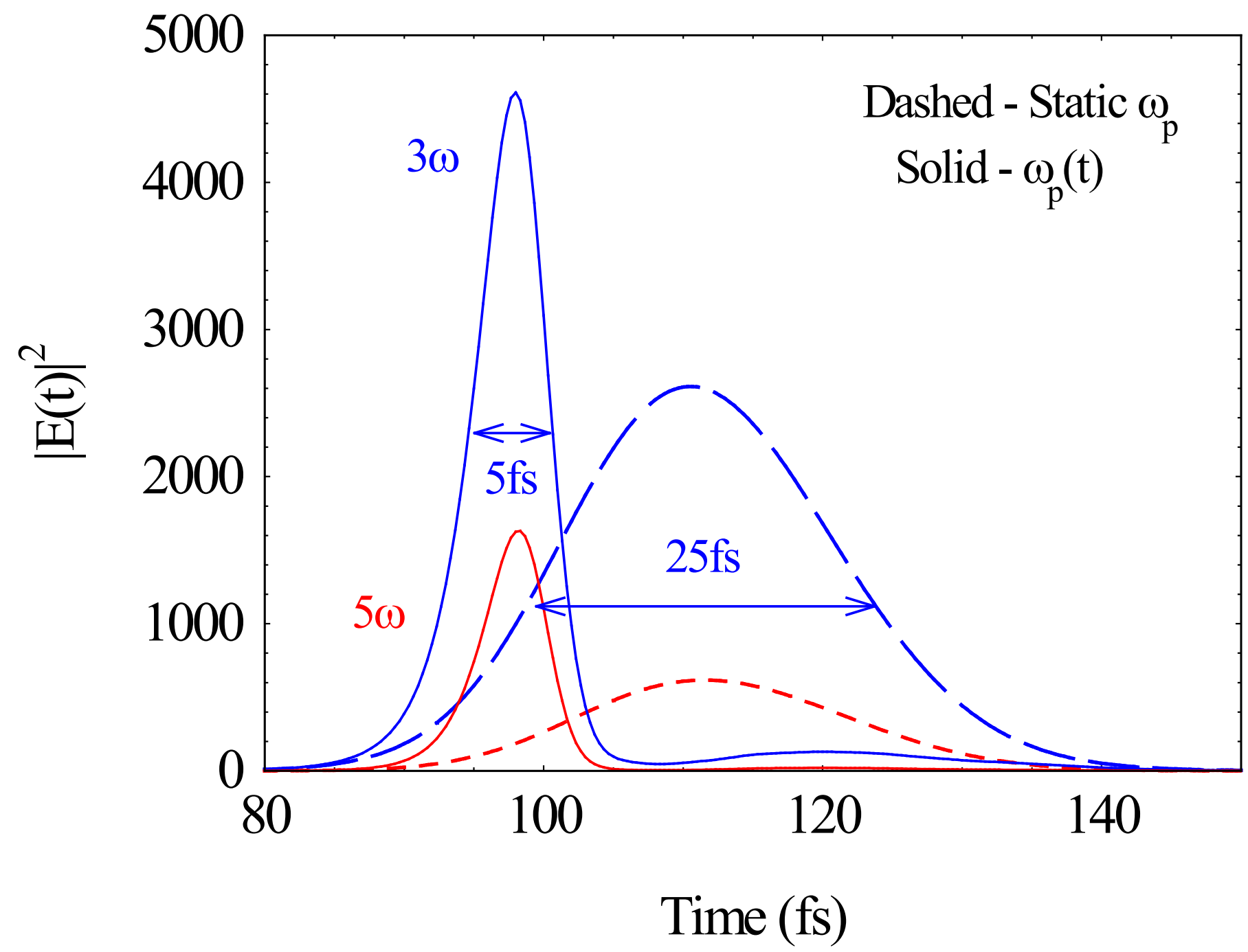

**Fig.3:** A ~43 fs, 190 GW/cm$^2$ pump pulse generates transmitted ~25 fs third harmonic (dashed blue curve) and 5th harmonic (dashed red curve). Modulating the plasma frequency as shown in Fig. 1, with a ~12 fs strong pump, reduces duration of the generated pulses to about 5 fs (solid blue and red curves, as indicated in the figure). Therefore, a broadened bandwidth is accompanied by significant pulse compression. The 7th harmonic undergoes similar dynamics.

The third predicted effect is *line splitting:* the plasma frequency doesn't jump instantly but rises and falls around the pump peak. Harmonic emission produced on the leading edge of the pump thus "feels" a different $\omega_p$ than emission on the trailing edge. In the frequency domain this can show up as two distinct peaks, which may be offset from the original harmonic frequency. The stronger the modulation depth (here 20%), the more pronounced the bifurcation between those two spectral lobes will be.

The fourth predicted effect is an overall redshift: as $\omega_p(t)$ changes in time reaching its minimum value when the pump pulse peaks inside the material, the material's ENZ point gradually shifts to longer wavelengths reaching its maximum value before recovering as the pulse exits the medium [40]. All harmonics generated during that interval inherit this bias, so the center of gravity of each spectral feature moves toward lower frequencies compared to the static-$\omega_p$ case. Taken together, these four signatures reveal a fundamental distinction between harmonic generation in stationary and time-varying media. In the former, the medium determines the harmonic spectrum.

In the latter, the evolving medium actively reshapes that spectrum during its generation. Harmonic emission therefore contains direct information about the plasma trajectory that produced it.

By contrast, with a fixed plasma frequency the refractive index undergoes minimal changes throughout the nonlinear interaction. Each harmonic then emerges from a single, unmodulated phase-matching condition, producing narrow, Lorentzian-like spectral lines with minimal chirp, no double-peaked structure, and no net wavelength shift. An exception in Fig. 2e is the transmitted spectrum of the unmodulated 6$^{th}$ harmonic, which is far less intense compared to all other harmonics, and shows broadening and multiple peaks around the center wavelength. Given its relative weakness compared to all other harmonics, a possible explanation for this behavior may be related to the fact that the 6$^{th}$ harmonic is driven by a number of source terms that combine harmonics with different phase matching conditions, and whose spectra may be only slightly modulated.

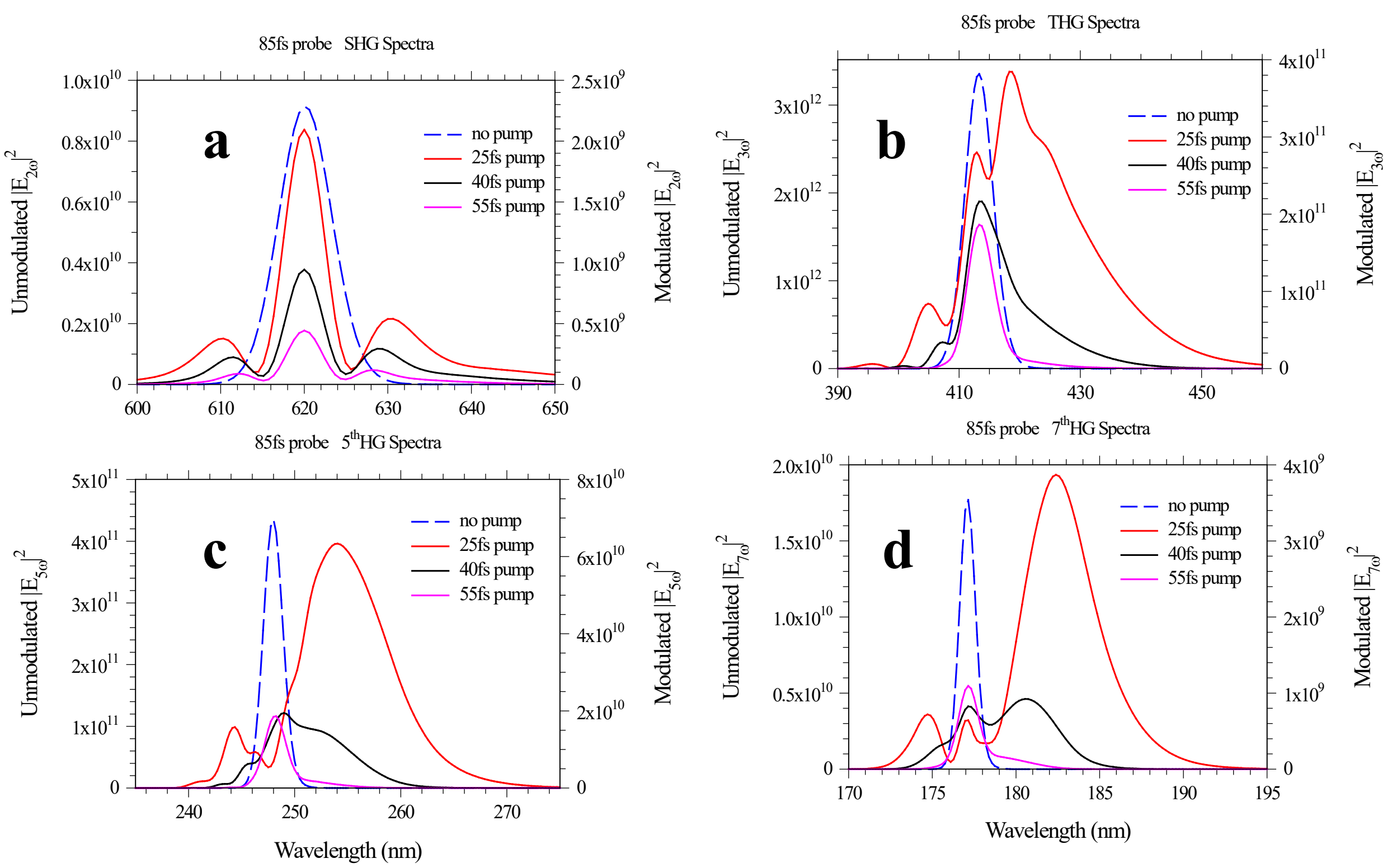

**Fig4:** Same as Fig. 2, for a ~85 fs, 190 GW/cm$^2$ control wave pulse, and 2-3 TW/cm$^2$ pump whose duration is as indicated in the figure. Only transmitted spectra are shown for: (a) SHG; (b) THG; (c) 5$^{th}$ HG; (d) 7$^{th}$ HG.

In Fig.4 we extend our predictions when the control wave is 85 fs in duration, while pump duration is as depicted in each of the captions. This figure demonstrates that the time-varying nature of the interaction persists even for control wave pulses of extended duration; it further indicates that spectral broadening and its associated effects are most efficiently induced when the pump pulse is shorter than the control wave. To conclude this section, we note that the same hydrodynamic-Maxwell framework can describe scenarios in which the plasma frequency rises on ultrafast timescales, instead of decreasing. By tuning the strong pump at or near a Lorentzian resonance [41-43] associated with an interband transition, one can promote electrons from the valence to the conduction band, thereby increasing the free-carrier density and driving a transient increase of $\omega_p$.

**Experimental validation**

To demonstrate the remarkably robust predictive capability of our hydrodynamic-Maxwell framework under real-world, large-signal conditions, we present experimental confirmation of these dynamics using a non-optimized, 240 nm-thick ITO layer under non-degenerate pump-control wave conditions (787 nm pump, 1250 nm control wave), both having pulse durations of approximately 100fs and 85 fs, respectively. Experimental setup and further details are outlined in the Supplementary Material. Although neither film thickness nor pulse durations were tailored for idealized harmonic conversion efficiencies, spectral broadening, or pulse compression, the resulting side-by-side transmitted spectra in Fig. 5 reveal a spectacular, fundamental behavioral split between even- and odd-order harmonics that validates the core architecture of our model.

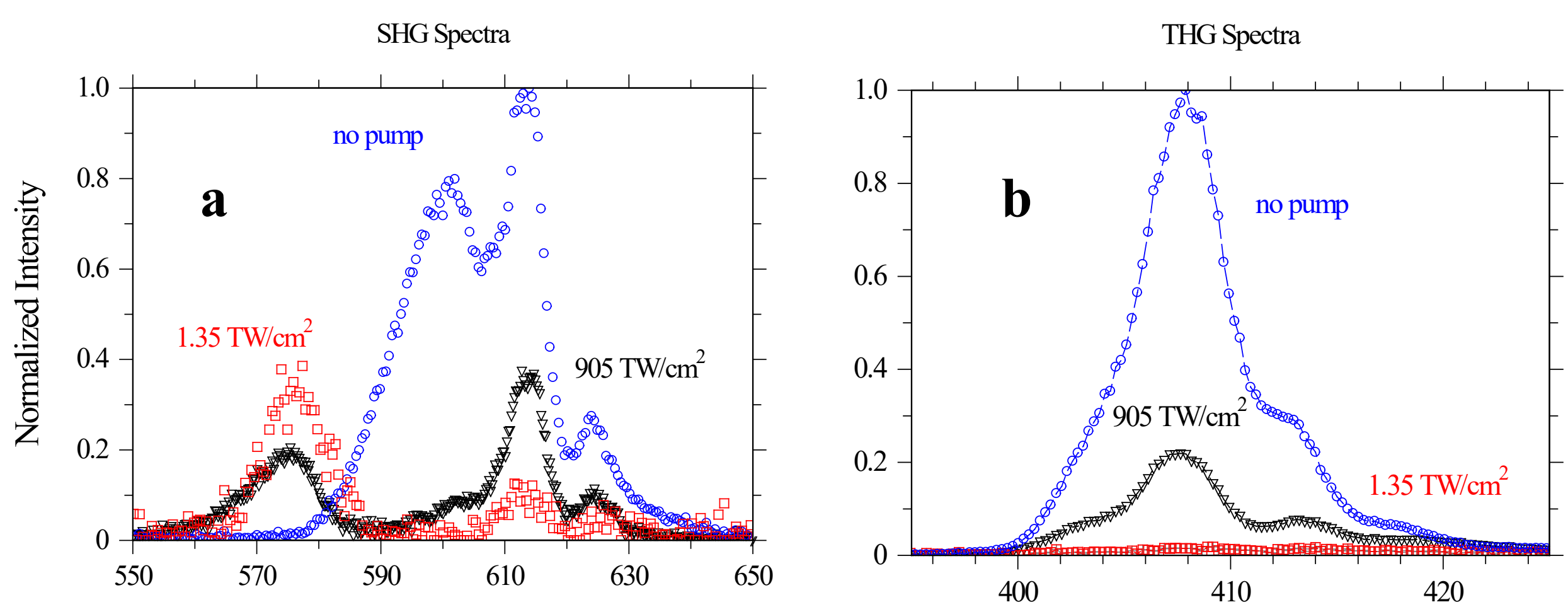

**Fig5:** Measured, normalized transmitted (a) second and (b) third harmonic generation spectra generated by a probe pulse ~85 fs, 150 GW/cm$^2$ tuned to 1250nm incident at 50° on a 240nm-thick ITO slab. The probe co-propagates with a ~100 fs, pump of variable peak power density, tuned to 787 nm and incident at 45°.

As shown in Fig. 5(b), the odd-order TH signal centered near 408 nm remains essentially unchanged at lower pump power densities, undergoes uniform amplitude quenching as peak power density is increased and is almost completely absorbed when the pump reaches 1.35 TW/cm². This behavior is a macroscopic signature of a bulk-conduction-driven process: within the hydrodynamic framework, extreme pump fields raise the internal electronic temperature, increase nonparabolicity, and drive a sharp rise in the scattering rate $\gamma$, in this case mostly through an intrinsic time dependence of the effective mass, as we will show in the Supplementary Material. Because bulk THG relies on long-lived, coherent anharmonic trajectories of the free-electron polarization, this rapid increase in dissipation suppresses the collective bulk response, producing the observed monotonic quenching.

In contrast, the even-order SH signal in Fig. 5(a) diminishes as it undergoes structural transformation. The multi-lobed baseline the signal displays when no pump is present collapses and is replaced by a dominant, blue-shifted channel. This divergence between odd- and even-order harmonics reflects the fundamentally different microscopic pathways involved: while odd-order THG is bulk-dominated and therefore highly sensitive to pump-induced damping, even-order SHG originates from surface Coulomb gradients, convective currents, and magnetic Lorentz forces, which respond instead to the instantaneous reshaping of the interfacial plasma. The importance of this distinction is that the experiment does not merely reveal different harmonic efficiencies. It reveals different microscopic pathways responding to the same evolving plasma environment.

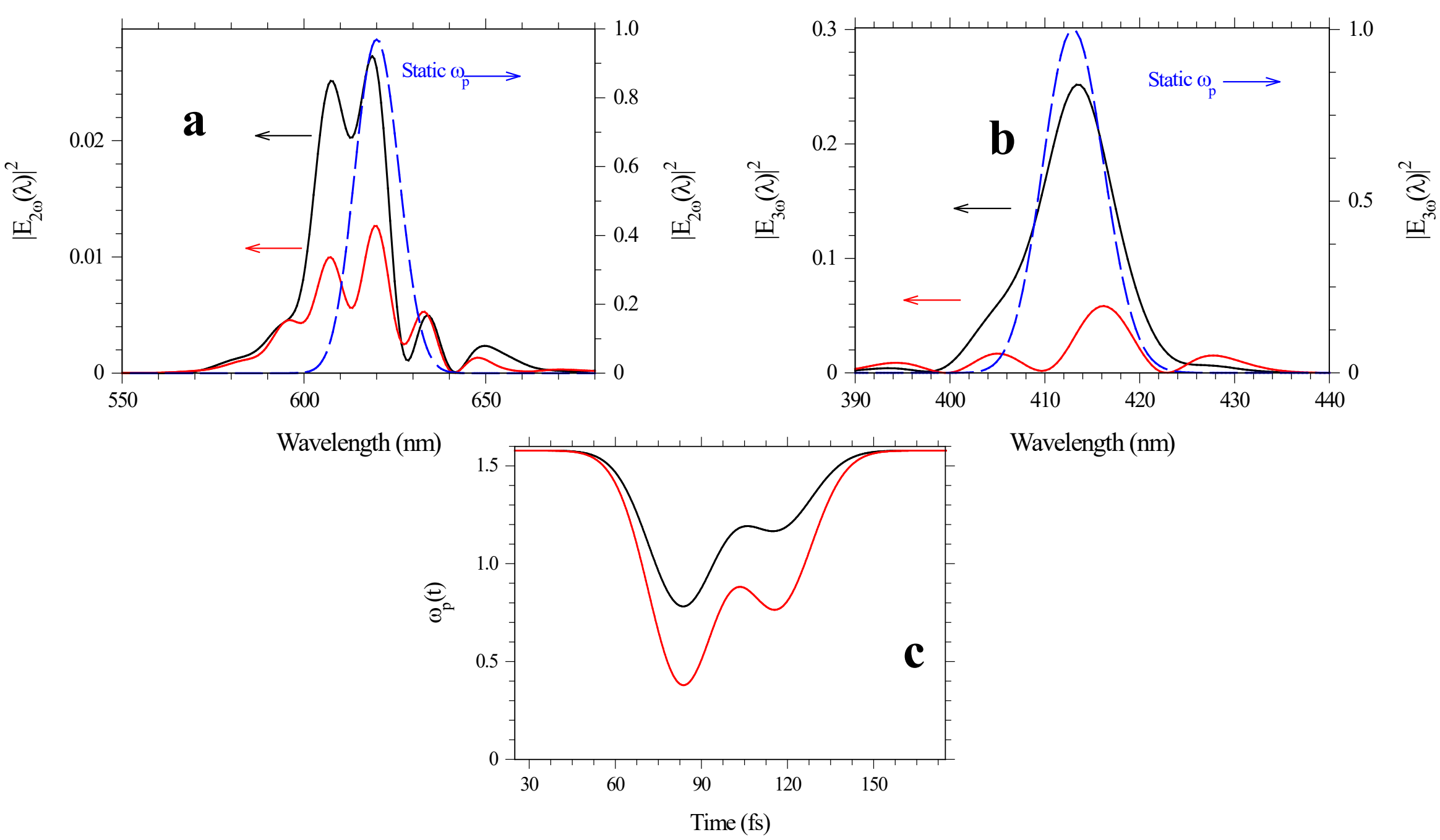

**Fig.6:** Simulated (a) second and (b) third harmonic spectra for similar pulse durations, tuning, peak power densities and angles of incidence. (c) temporal evolution of the plasma frequencies used to calculate the modulated spectra in (a) and (b).

Figures 6(a) and (6b) present simulated SH and TH spectra for comparable angles, pump intensities, and tuning, for the timed-dependent plasma frequencies shown in Fig.(6c). The pump-probe portion of the simulations reveals that at oblique incidence and high intensities the temporal profile of the plasma frequency in a thick layer does not follow a single inverted Gaussian. Instead, at moderately high intensities it acquires at least two distinct minima, generally forming an inverted double-peaked Gaussian envelope, as shown in Fig.6c, which is the function we used in the simulation. This behavior is consistent with earlier findings in [39], where a 300 nm ITO layer exhibited the same dual-minimum structure. The origin is a complex interplay of geometry, spatio-temporal dispersion, and spatial inhomogeneities in the pump and probe fields. These fields acquire longitudinal components that drive the free-electron fluid to oscillate within the layer, forming standing-wave patterns that produce nonuniform excitation and a correspondingly intricate temporal evolution of the plasma frequency. In the Supplementary Material we expand on this point and show the full complexity of the plasma frequency for 240 nm-thick layers at oblique incidence and address the appearance of sidebands in the no-pump scenario that emerges in Fig.5. The same qualitative dynamics appear in the 25 nm-thick layers, although in that case geometric effects are largely suppressed. While in the simulation the central SH peak is diminished but survives, more complex temporal profiles of the plasma frequency like those shown in the Supplementary Material may suffice to more closely align simulation's results with experimental findings.

Non-monotonic, ultrafast modulations of the plasma frequency in thick ITO films under degenerate pump-probe excitation has been reported recently in reference [42]. In that case the authors attributed the dynamics to a competition between intraband heating and delayed Auger-type interband excitation. While their extended two-temperature model captures the thermodynamic evolution of $\omega_p(t)$ and $\gamma(t)$ in a linear regime, the physical origin of the non-monotonicity differs fundamentally and substantially from the electrodynamic mechanisms identified in our hydrodynamic-Maxwell framework. In our case, the double-minimum structure of $\omega_p(t)$ that we use in our simulations is triggered by temperature increase of the free electron gas but is also strongly affected by pump-probe geometry, internal interference, longitudinal field components that drive free electron gas oscillations, spatial inhomogeneities, linear and nonlinear spatio-temporal dispersions that determine standing-wave patterns inside the film. Thus, although both approaches observe qualitatively similar departures from a single inverted Gaussian profile, they operate in distinct regimes and describe complementary physical processes: in reference [42] the authors reconstruct carrier-scattering dynamics from Fresnel-coefficient retrieval, whereas our model predicts the full microscopic electrodynamics governing nonlinear frequency conversion and harmonic generation in time-varying media, thus allowing for a reliable ultra-broad band analysis of higher order harmonics.

These observations have critical implications. Because the plasma-frequency trajectory depends on accurate knowledge of parameters like effective masses, carrier densities, and linear and nonlinear dispersion, it is extremely difficult to infer the true experimental plasma frequency modulation directly from the measured harmonic spectra alone. Many distinct temporal profiles can produce qualitatively similar spectral outcomes, especially since geometric resonances and field-distribution inhomogeneities are unavoidable in thick samples. The inverted double-peaked Gaussian used in our simulations should therefore be viewed not as a literal reconstruction of the experiment, but as a physically motivated representative of the broad family of temporal modulations compatible with the data.

In this sense, the agreement between experiment and simulation does not imply that the plasma frequency is known precisely; rather, it demonstrates that the hydrodynamic-Maxwell framework captures the correct microscopic mechanisms governing the interaction. The model identifies the qualitative signatures, namely bulk quenching of odd harmonics, interfacial reshaping of even harmonics, spectral shifts, and structural reconfiguration, that arise naturally from time-dependent plasma dynamics, even when the exact temporal profile cannot be uniquely determined.

This behavioral divergence highlights the distinct role of the time-varying interface. Because even-order SHG is strictly a manifestation of spatial symmetry-breaking, driven at the surface via Coulomb gradients, convective currents, and magnetic Lorentz forces, it is not merely suppressed by increased bulk damping. Instead, the rapid, pump-induced sweep of the instantaneous plasma frequency dynamically modulates the phase-matching condition and intensifies field confinement at the interface as the peak of the pulse reaches the ITO layer. This forces the emerging nonlinear radiation into a dynamically blue-shifted emission pathway. By capturing this clear experimental dichotomy between bulk-quenched and surface-reconfigured harmonic channels under identical macroscopic excitation states, these findings confirm that the independent microscopic pathways on display in our hydrodynamic fluid model are robustly exhibited in our experiment, even in settings that are far from optimal.

**Summary and Concluding Remarks**

In this work, we have demonstrated a two-field approach for controlling high-order harmonic emission through the ultrafast evolution of the material itself. An intense optical field generates its own broadband high-harmonic comb as it modifies the electronic plasma and consequently reshapes the individual harmonic lines generated by a control field. This mechanism is fundamentally distinct from wave-mixing-based frequency conversion. Rather than generating new spectral components through frequency- and wave-vector-dependent mixing channels, it enables the simultaneous modification of multiple collinear harmonics through a time-dependent material response. The deeply subwavelength nature of the interaction further removes the conventional constraints associated with propagation and phase matching, providing a particularly flexible platform for controlling the harmonic spectrum.

The resulting control extends well beyond a simple shift of the harmonic spectrum. Ultrafast changes in carrier density and effective mass modify the plasma frequency and dynamically alter the refractive index experienced by the generated harmonics. Our simulations reveal spectral broadening, line splitting, frequency shifts, and substantial pulse compression across multiple harmonic orders. Importantly, individual harmonic channels can respond differently to the evolving material, with their spectral modification depending on parameters including the intensity, polarization, and temporal properties of the interacting fields. The time-dependent plasma therefore provides multiple degrees of freedom for simultaneously manipulating different components of the harmonic comb, introducing a route towards broadband hyperspectral shaping of nonlinear emission.

The self-consistent hydrodynamic-Maxwell description provides the microscopic interpretation of these effects. By combining free-electron hydrodynamics including pressure, viscosity, magnetic Lorentz forces, convective nonlinearities, nonlocality, and hot-electron effects with Lorentz oscillators for bound charges, both even- and odd-order harmonics naturally emerge from their underlying microscopic sources without the need to invoke phenomenological nonlinearities [32]. Surface Coulomb, magnetic-dipole, convective, and nonlocal contributions account for symmetry-breaking even-order emission, while free- and bound-electron dynamics determine the higher odd-order response. The same description shows that pump-induced changes in effective mass and carrier density introduce additional damping mechanisms that modify the coherence time of the nonlinear polarization [19, 39, 43]. It is also important to recognize that the notion of damping itself may require revision in the strong-field regime. Earlier work based on an alternative radiation-reaction formalism demonstrated that replacing the traditional Abraham-Lorentz description introduces a hierarchy of additional linear, nonlinear, and nonlocal damping terms that naturally couple to charge accumulation, magnetic-field gradients, and other electrodynamic quantities [44]. Under ordinary conditions these corrections are small; however, in the TW/cm² regime considered here, where effective masses, carrier densities, and plasma frequencies evolve dynamically, such terms may contribute to the overall reshaping of harmonic emission and should not be discarded a priori.

Experiments performed on ITO films provide experimental validation of the physical picture predicted by the model and reveal the markedly different response of bulk-driven odd harmonics and interface-driven even harmonics under strong pumping. Together with the simulations, the experiments also identify the critical conditions for maximizing spectral reshaping: strongly asymmetric pulse durations, deeply subwavelength films, and excitation intensities approaching the TW/cm² regime. These results establish time-dependent plasma dynamics as a powerful new tool for broadband hyperspectral shaping of multiple harmonic channels beyond the framework of four-wave-mixing processes. Our findings are therefore of particular relevance to ultrafast science, opening opportunities for ultrafast pulse manipulation and the generation and control of tailored multi-harmonic waveforms. More broadly, these results suggest that harmonic generation in strongly driven conductors should not always be viewed as a process taking place within a predefined material response. Under sufficiently intense excitation, the material response itself becomes a dynamic degree of freedom that participates directly in the frequency-conversion process.

**Supplementary Material**

# Plasma and light: Shaping High Harmonic Generation in Time-Varying Media

W. Jaffray[1], M. Ferrera[1], M. A. Vincenti[2], D. de Ceglia[2], C. Rizza[3],
N. Litchinitser[4], J. Trull[5], and C. Cojocaru[5], M. Scalora[1]

[1]Institute of Photonics and Quantum Sciences, Heriot-Watt University, SUPA, Edinburgh EH144AS, United Kingdom
[2]Department of Information Engineering, University of Brescia, 25123 Brescia, Italy
[3]Department of Physical and Chemical Sciences, University of L'Aquila, I-67100 L'Aquila, Italy
[4]Department of Electrical and Computer Engineering, Duke University, Durham, North Carolina 27708, USA
[5]Department of Physics, Universitat Politècnica de Catalunya, 08222 Terrassa (Barcelona), Spain

The simplified nonlinear oscillator model used in reference [32] of the main text inaccurately blends multiple physical effects into a single effective nonlinearity, obscuring: (1) Surface (Coulomb) nonlinearities arising at interfaces via $(\nabla\cdot\mathbf{P})\mathbf{E}$ terms for free carriers, and $(\mathbf{P}\cdot\nabla)\mathbf{E}$ for bound charges. SHG from aluminum films provides a dramatic example of this [R1]; (2) Magnetic-dipole contributions from the Lorentz force term $\mathbf{J}\times\mathbf{B}$; (3) Convective hydrodynamic terms like $(\mathbf{J}\cdot\nabla)\mathbf{J}$ and $(\nabla\cdot\mathbf{J})\mathbf{J}$ in the free-electron fluid; (4) Linear nonlocal pressure and viscosity terms, $\nabla(\nabla\cdot\mathbf{P})$ and $\nabla^2\mathbf{P}$, respectively, that impart *k*-dependent dispersion; (5) Nonlinear nonlocal quantum contributions proportional to $(\nabla\cdot\mathbf{P})\nabla(\nabla\cdot\mathbf{P})$ that may become significant in heavily doped semiconductors [R2]; (6) Temperature (absorption)-driven plasma-frequency shifts due to hot-electron dynamics; (7) Bound-electron resonances contributing both linear dispersion and intrinsic Lorentzian nonlinearity; and (8) As we will see below, additional nonlinear damping terms arise under fast response dynamics from time-dependent effective mass and charge density [R3-R5].

Traditional methods to study harmonic generation from centrosymmetric metal surfaces are based on the free electron model and have consistently attempted to define effective surface (longitudinal and tangential) and volume second order nonlinearities [R6]. These effective $\chi^{(2)}$ are defined in terms of *a*, *b*, and *d* coefficients, which are typically of order unity:

$$\chi^{(2)}_{\perp,surface}=a(\omega)\frac{\varepsilon(\omega)-1}{16\pi m\omega^2};\qquad \chi^{(2)}_{\|,surface}=b(\omega)\frac{\varepsilon(\omega)-1}{16\pi m\omega^2};\qquad \chi^{(2)}_{bulk}=d(\omega)\frac{\varepsilon(\omega)-1}{16\pi m\omega^2}\quad, \qquad \text{(S1)}$$

where $\varepsilon(\omega)$ refers to the real part of the dielectric constant and describes strictly a Drude-like material response. Eqs.S1 are derived in a context of thick metal layers, which allow for minimal penetration depth, and are not applicable to thin, semitransparent layers. In that case, spatial derivatives like $\nabla\cdot\mathbf{P}$, which define the spatial charge distribution, are non-zero away from the surface, further complicating the picture of what is generated at the surface and what originates within the volume, making this a pursuit with limited prospects. In addition, and perhaps critically, the free electron model fails spectacularly near resonances, as it does for aluminum [R1]. It can also be demonstrated that nonlocal blueshifts of ENZ resonances and hot-electron-induced redshifts of the plasma frequency crucially modify both SHG and THG yields in thin ITO layers [R7]. Recent ENZ-SHG and high harmonic generation (HHG) [R8] also underscores the fact that preserving full linear and nonlinear dispersions is essential for predictive modeling of harmonic conversion with a pump tuned near the ENZ point, and harmonics generated deep in the UV range and beyond. We thus proceed using a rigorous hydrodynamic-Maxwell framework, which couples Maxwell's equations to a two-component medium: a hydrodynamic description of free electrons plus Lorentz oscillators for bound charges, automatically decomposing SH, TH, and higher order sources into Coulomb, magnetic, convective, and pressure terms.

**Tackling The Two Pump Problem**

High harmonic generation transforms intense laser fields into signals spanning deep ultraviolet to soft X-ray wavelengths. Traditionally driven by a single ultrafast laser [R9-R12], we explore HHG when two pumps interact within the same medium. In our scheme, an ultrafast, intense pump pulse co-propagates with a much weaker control wave. While each field produces its own train of harmonics, the pump's rapid modulation imprints the control wave with a time-varying plasma frequency [R5] that can significantly reshape the harmonics generated by the control wave compared to single incident pulse scenarios [R8]. This interplay opens new vistas to tunable attosecond sources, dynamic spectral control, and injects fresh insights into plasma nonlinearity. In our approach the interaction thus unfolds within the hydrodynamic-Maxwell framework as described in references [R5] and [R8]. As the pump plows into the medium, the evolving plasma modifies the dielectric response experienced by both pump and control wave, creating a highly dynamic backdrop for the nonlinear spectrum generated by the control wave. Unlike simpler cold-plasma descriptions (an electron gas at equilibrium, with no thermal motion, no pressure, no viscosity, no hot-electron population, and no spatial dispersion), the hydrodynamic-Maxwell approach faithfully reproduces the rapid onset and evolution of spatio-temporal dynamics that ensues in transparent conductive oxides under extreme conditions [R5], which include few-cycle pulses, and $TW/cm^2$ peak power densities [R13].

**Method of Solution: Step One**

In addition to a rigorous theoretical framework, crafting clear, disentangled spectra from two pumps also demands careful consideration of the geometry. For a degenerate configuration, when both pumps share the same carrier wavelength, they should be incident at different angles. This angular separation translates each pump's harmonics into distinct spatial channels, preventing spectral overlap. Upon exiting the medium, the beams are spatially separated along their unique directions, enabling straightforward isolation of each set of harmonic signals.

In the non-degenerate configuration, using different pump wavelengths naturally produces interweaved harmonic spectra. However, if the pump and control wave beams are incident collinearly, the nondegenerate interaction can also generate FWM signals at frequencies that could overlap with the harmonic spectra. Potentially, these additional signals can contaminate the control wave's measured spectra. This can be avoided by introducing a small angle between pump and control wave beams, so that the generated FWM signal is emitted in a different direction. We note that the generation of even harmonics in flat centrosymmetric media requires oblique incidence. Finally, the degree of synchronization between the two pulses can affect both the spectral characteristics of the control wave [R5, R13] and its generated harmonics [R8]. Significantly, the spectral modifications (broadening, line splitting, red-shifting, and pulse compression) predicted by this framework are distinct from standard parametric third-order processes, such as instantaneous XPM or FWM. While a traditional $\chi^{(3)}$ description relies on instantaneous, virtual-state transitions within a static electronic backdrop, our time-varying approach captures a non-adiabatic, real-carrier path redistribution. The pump pulse actively renormalizes the intrinsic Drude parameters of the conduction band plasma on a femtosecond timescale. This creates non-equilibrium material trajectories and dynamic damping contributions (see below) that standard parametric wave-mixing tensors fundamentally cannot encapsulate. We thus set out to unravel these dynamics where the pump propagates unaffected by the control wave, but where the control wave and its harmonics are modified by the interaction.

The order-of-magnitude discrepancy in peak power densities between the pump and control wave justifies a standard control wave-field approximation, enabling a two-step solution to the problem. In step one, what occurs between the two beams may be treated separately from harmonic generation as a *pump-control wave* interaction, where we focus on the extraction of the time dependent plasma frequency $\omega_p(\mathbf{r},t) = \left( \frac{4\pi n_f(\mathbf{r},t) e^2}{m_f^*(\mathbf{r},t)} \right)^{1/2}$ experienced by the control wave [R5], while retaining full nonlinear dynamics for the pump. $m_f^*(\mathbf{r},t) \approx m_0^* + \Lambda \iint \mathbf{J} \bullet \mathbf{E} d\mathbf{r}^3 dt$ is the effective mass, with a constant of proportionality $\Lambda$ ; $m_0^*$ is the electron's rest mass; $\Lambda \iint \mathbf{J} \bullet \mathbf{E} d\mathbf{r}^3 dt$ represents absorption; $\mathbf{J} = \dot{\mathbf{P}}_f$ is the free current density, $\mathbf{P}_f$ is the corresponding polarization component; $n_f(\mathbf{r},t) = n_{0f} - \frac{1}{e} \nabla \bullet \mathbf{P}_f$ is the free electron density and $n_{0f}$ is the background charge density with no applied field; and $\mathbf{E}$ is the electric field. The details of the theoretical aspects of pump-control wave dynamics in extreme regimes are presented in [R5]. Suffice it to say here that the control wave intensity is chosen to be sufficiently weak (of order 100 GW/cm$^2$; in this case our simulations suggest that in this intensity range the control wave can modulate its own plasma frequency by approximately only 1%,) that it is possible to extract a *simplified time-dependent, instantaneous plasma frequency* as given by the expression [R5]:

$$\omega_p(\mathbf{r},t) = \left( \omega_{p,0}^2 - \tilde{\Lambda} \,|\, \mathbf{E}_{pump} \,|^2 + \tilde{\Lambda}^2 \,|\, \mathbf{E}_{pump} \,|^4 - \tilde{\Lambda}^3 \,|\, \mathbf{E}_{pump} \,|^6 + \tilde{\Lambda}^4 \,|\, \mathbf{E}_{pump} \,|^8 \right)^{1/2}, \quad \text{(S2)}$$

where $\tilde{\Lambda} = \Lambda \sigma_0 V \tau_0 / m_0^*$ is a coefficient scaled by the interaction volume $V$; the temporal duration of the pulse $\tau_0$; and the frequency dependent conductivity $\sigma_0$. We emphasize that in this portion of the dynamics there are additional terms that couple pump and control fields that may not be neglected if intensities of the two beams are of the same order of magnitude, or if oblique incidence is combined with circular and/or elliptical polarization. These terms can introduce effective anisotropic and magneto-optics effects enriching light dynamics [R3] that would require a different, more extensive approach. As a result, we proceed with the understanding that Eq. S2 is valid in the regime in which we operate.

**Method of Solution: Step Two**

For completeness, we briefly summarize the structure of the material equations. The free-electron polarization includes surface Coulomb terms, magnetic Lorentz forces, convective nonlinearities, nonlocal pressure and viscosity, and hot-electron corrections, while the bound-electron oscillators contribute both surface and intrinsic bulk nonlinearities. These contributions are written explicitly in Eqs. S3-S6. The classical free and bound electron current densities are defined as $\mathbf{J}_f = \dot{\mathbf{P}}_f$, and $\mathbf{J}_{bj} = \dot{\mathbf{P}}_{bj}$ , respectively. The insertion of a time dependent plasma frequency in the control wave's equation of motion leads to the following scaled equations for conduction and core electrons:

$$\ddot{\mathbf{P}}_f + \tilde{\gamma}_f \dot{\mathbf{P}}_f = \frac{n_{0f} e^2 \lambda_r^2}{m_f^*(\mathbf{r},t) c^2} \mathbf{E} + \frac{e\,\lambda_r}{m_f^*(\mathbf{r},t) c^2} \left(\nabla \bullet \mathbf{P}_f\right) \mathbf{E} + \frac{e\,\lambda_r}{m_f^*(\mathbf{r},t) c^2} \dot{\mathbf{P}}_f \times \mathbf{H}$$

$$- \frac{1}{n_{0f} e \lambda_r} \left[ \left(\nabla \bullet \dot{\mathbf{P}}_f\right) \dot{\mathbf{P}}_f + \left(\dot{\mathbf{P}}_f \bullet \nabla\right) \dot{\mathbf{P}}_f \right] + \frac{3}{5} \frac{E_F}{m_f^*(\mathbf{r},t) c^2} \left[ \nabla \left(\nabla \bullet \mathbf{P}_f\right) + \frac{1}{2} \nabla^2 \mathbf{P}_f \right] \quad \text{(S3)}$$

$$\ddot{\mathbf{P}}_{bj} + \tilde{\gamma}_{bj} \dot{\mathbf{P}}_{bj} + \tilde{\omega}_{0,bj}^2 \mathbf{P}_{bj} + \mathbf{P}_{bj}^{NL} = \frac{n_{0,b} e^2 \lambda_0^2}{m_{bj}^* c^2} \mathbf{E} + \frac{e\,\lambda_0}{m_{bj}^* c^2} (\mathbf{P}_{bj} \bullet \nabla) \mathbf{E} + \frac{e\,\lambda_0}{m_{bj}^* c^2} \dot{\mathbf{P}}_{bj} \times \mathbf{H} \quad , \quad \text{(S4)}$$

where

$$\frac{n_{0,f} e^2 \lambda_r^2}{m_f^*(\mathbf{r},t) c^2} \mathbf{E} = \pi \tilde{\omega}_p^2(t) \mathbf{E} - \tilde{\Lambda} \left(\mathbf{E} \bullet \mathbf{E}\right) \mathbf{E} + \tilde{\Lambda}^2 \left(\mathbf{E} \bullet \mathbf{E}\right)^2 \mathbf{E} - \tilde{\Lambda}^3 \left(\mathbf{E} \bullet \mathbf{E}\right)^3 \mathbf{E} \quad \text{(S5)}$$

and

$$\mathbf{P}_{bj}^{NL} = -\tilde{\beta}_{bj} (\mathbf{P}_{bj} \bullet \mathbf{P}_{bj}) \mathbf{P}_{bj} + \tilde{\delta}_{bj} (\mathbf{P}_{bj} \bullet \mathbf{P}_{bj})^2 \mathbf{P}_{bj} - \tilde{\theta}_{bj} (\mathbf{P}_{bj} \bullet \mathbf{P}_{bj})^3 \mathbf{P}_{bj} \quad \text{(S6)}$$

At this stage all fields are real. The frequency $\tilde{\omega}_p^2(t)$ in Eq. 5 is generated by the pump, while the electric, magnetic, and polarization fields depicted in Eqs. S2-S6 now refer to the control wave. The subscript *b* in Eqs. S4 and Eq. S6 stands for *bound*, and the counter *j* represents the *j*^th bound oscillator species. The linear, local dielectric response of ITO is modelled with one Drude component and one Lorentzian function resonant in the 200-300nm range (see Fig.1b in the main text):

$$\varepsilon(\omega) = 1 - \frac{\tilde{\omega}_{pf}^2}{\omega^2 + i\tilde{\gamma}_f \omega} - \frac{\tilde{\omega}_{p1}^2}{\omega^2 - \tilde{\omega}_{01}^2 + i\tilde{\gamma}_{01}\omega}, \quad \text{(S7)}$$

where $\left(\tilde{\omega}_{pf}, \tilde{\omega}_{p1}\right) = \left(1.5785, 11.5\right)$, $\left(\tilde{\gamma}_f, \tilde{\gamma}_{01}\right) = \left(0.1105, 0.01\right)$, and $\left(\tilde{\omega}_{0f}, \tilde{\omega}_{01}\right) = \left(0, 7\right)$. Eq. (S7) is obtained by Fourier transforming Eqs. S3-S4 after dropping all nonlinear and nonlocal terms. The angular frequency $\omega$ is normalized to $\omega_{norm} = 2\pi/1\mu m$. Eqs. S3-S4 include descriptions of a time dependent plasma frequency $\tilde{\omega}_p^2(t)$, surface, $\left(\nabla \bullet \mathbf{P}_f\right)\mathbf{E}$, magnetic, $\dot{\mathbf{P}}_f \times \mathbf{H}$, convective $\left[\left(\nabla \bullet \dot{\mathbf{P}}_f\right)\dot{\mathbf{P}}_f + \left(\dot{\mathbf{P}}_f \bullet \nabla\right)\dot{\mathbf{P}}_f\right]$, nonlocal $\left(\nabla\left(\nabla \bullet \mathbf{P}_f\right) + \frac{1}{2}\nabla^2 \mathbf{P}_f\right)$ and hot electron $(\sim \tilde{\Lambda}\left(\mathbf{E} \bullet \mathbf{E}\right)\mathbf{E} + \ldots)$ nonlinearities in the free electron polarization component. Similarly, Eq.(S4) also contains surface, $(\mathbf{P}_{bj} \bullet \nabla)\mathbf{E}$, magnetic, $\dot{\mathbf{P}}_{bj} \times \mathbf{H}$, and bulk $-\tilde{\beta}_j (\mathbf{P}_{bj} \bullet \mathbf{P}_{bj})\mathbf{P}_{bj} + \tilde{\delta}_j (\mathbf{P}_{bj} \bullet \mathbf{P}_{bj})^2 \mathbf{P}_{bj} - \tilde{\theta}_j (\mathbf{P}_{bj} \bullet \mathbf{P}_{bj})^3 \mathbf{P}_{bj}$ nonlinearities triggered by bound electrons [R1, R14]. We neglect even order nonlinearities in centrosymmetric systems and expand the fields, and free and bound electron nonlinearities to seventh order to account for up to 7[th] harmonic generation. The nonlinear coefficients from bound electrons can be derived directly from a classical nonlinear oscillator model and may be written as $\tilde{\beta}_j = \omega_{0,j}^2 \lambda_0^2 / \left( L^2 n_{0b}^2 e^2 c^2 \right)$, and $\tilde{\delta}_j = \tilde{\beta}_j / n_{0b}^2 e^2 L^2$ $\tilde{\theta}_j = \tilde{\beta}_j / n_{0b}^4 e^4 L^4$ . These coefficients are real, known, and depend only on linear oscillator parameters like resonance frequency $\omega_{0,j}$ and lattice constant *L*. $E_f$ is the Fermi energy; $\lambda_0 = 1\mu m$ is a conveniently chosen scaling wavelength; *c* is the speed of light in vacuum; $m_0^*$ and $m_{bj}^*$ are the free and bound electron effective masses, respectively. $\xi = z / \lambda_0$, $\varsigma = y / \lambda_0$, $\zeta = x / \lambda_0$, $\tau = ct / \lambda_0$ are the scaled space and time coordinates that are used to calculate spatial and temporal derivatives. Damping coefficients, plasma and resonance frequencies are scaled as follows: $\tilde{\omega}_p^2 = \frac{4\pi ne^2}{m_0^*} \frac{\lambda_0^2}{c^2}$, $\tilde{\omega}_0 = \omega_0 \frac{\lambda_0}{c}$, $\tilde{\gamma} = \gamma \frac{\lambda_0}{c}$. Once the dielectric function data is fitted using the dielectric function in Eq.S7, the ratios $n_{0f} / m_0^*$ and $n_{0,b} / m_{bj}^*$ are set and suffice to allow the linear portion of the dynamics to evolve in both Eqs. S3 and S4. However, the nonlinear coefficients require direct knowledge of both $n_{0f}$ and $m_0^*$ in Eq.S3, and of $m_{bj}^*$ in Eq. S4. These coefficients determine the relative magnitudes of competing sources of even and odd harmonics: Coulomb, magnetic Lorentz, and convective terms drive even harmonic generation; hot electrons in the free electron current and bound charges determine the efficiency of odd harmonics. For details about model and method of solution, we refer the reader to references [R1, R3-R5, R7-R8, R14-R16]. The integrations are performed using a split-step method coded in Fortran, that advances the fields in free space using an unconditionally stable fast Fourier transform algorithm. The material equations are solved using a second-order accurate predictor-corrector method [R16]. Spatial and correspondingly causal temporal integration steps are δz = δy = δx =1 nm and δt = $3.3\times10^{-18}$ sec., respectively. The calculations are carried out on ordinary desktop computers, with execution times that range from a few minutes to several hours, depending on pulse duration and angle of incidence. However, we note that the code can also be optimized for complex two-dimensional gratings, maintaining similar integration steps and performance as exemplified in references [R17] and [R18]. Addressing one- and two-dimensional geometries is usually sufficient to extract the fundamental physics necessary to guide device design, as these dimensions capture the vast majority of practical nanostructured and plasmonic architectures without sacrificing computational tractability.

## Pump-Control wave Simulations and the Origin of the Double-Dipped Plasma Frequency Modulation

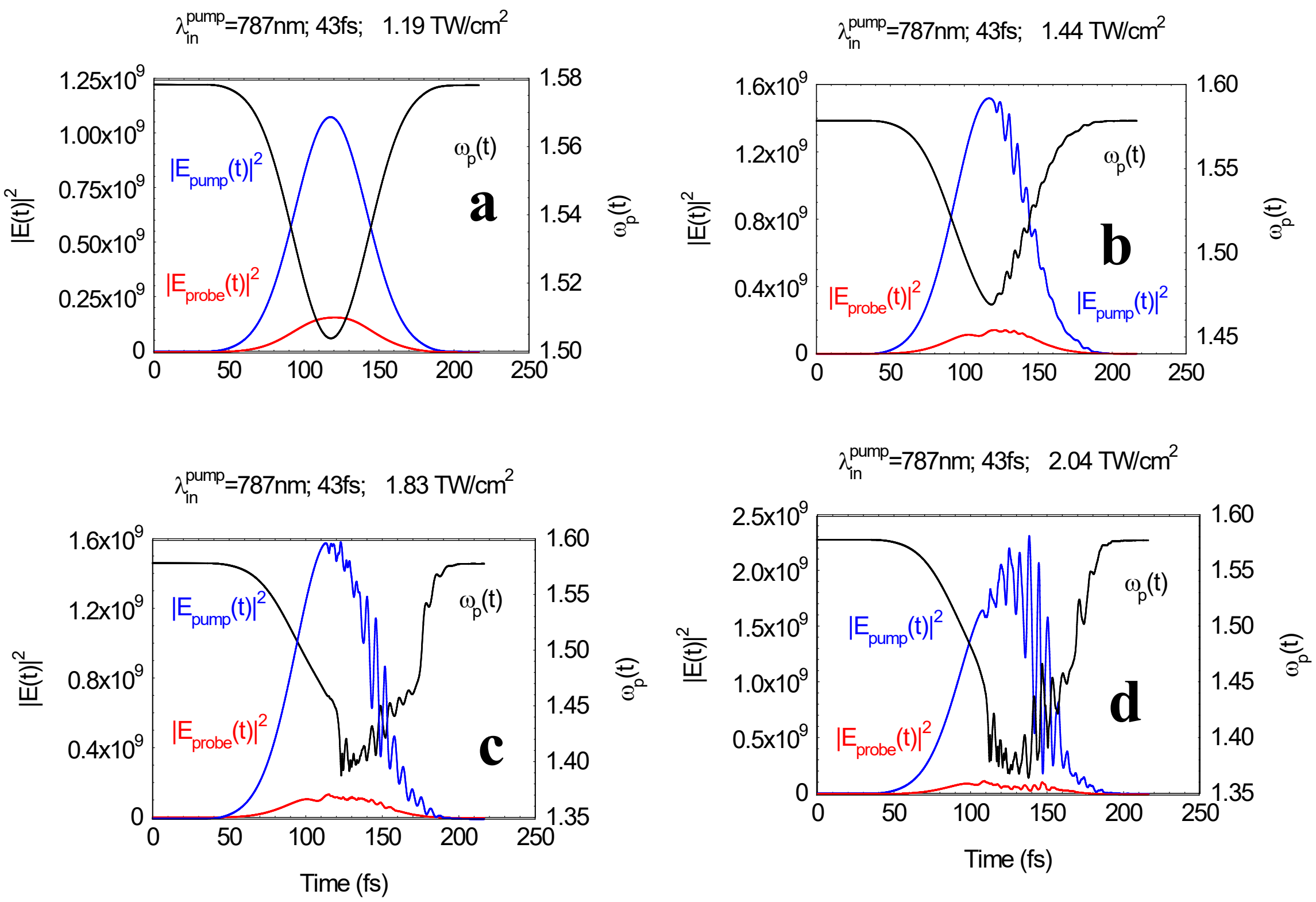

Fig. FS1: Pump and control wave field intensities, and plasma frequency as functions of increasing pump peak power density: (a) 1.19 TW/cm$^2$, (b) 1.7 TW/cm$^2$, (c) 1.83 TW/cm$^2$, (d) 2.04 TW/cm$^2$.

To justify the temporal modulation of the plasma frequency used in the main text simulations of harmonic generation, we performed a series of pump-control wave calculations in which the pump pulse was held fixed while the control wave was tuned to 1240 nm, near the ENZ crossing point. The four panels in Fig. FS1 display the resulting temporal evolution of $|E_{\text{pump}}(t)|^2$, $|E_{\text{control wave}}(t)|^2$, and the instantaneous plasma frequency, as the pump peak power density is progressively increased. At the lowest pump intensity in panel (a), $\omega_p(t)$ follows a single inverted Gaussian envelope that closely tracks the pump profile. As the pump power density increases, the temporal response of the plasma frequency becomes increasingly complex. Panel (b) shows the emergence of oscillatory behavior with secondary minima: the dynamics of conduction electrons no longer follow a simple monotonic excursion. In panel (c) the dips become more pronounced, reflecting the onset of geometric and interference effects inside the film. By panel (d), at the highest pump intensity, $\omega_p(t)$ develops several well-defined minima, forming a combination of an inverted Gaussian and sinusoidal structure. As the pump drives stronger internal fields, spatial inhomogeneities, standing-wave patterns, and nonuniform excitation across the film thickness produce multiple temporal extrema in the plasma frequency. These features are not numerical artifacts: they arise naturally from the hydrodynamic equations of motion when realistic pump-control wave geometries, linear and nonlinear dispersions, and boundary conditions are included. The four simulations therefore provide a clear physical justification for the simple, inverted double-dipped Gaussian modulation used in the main text section where harmonic generation was calculated. They thus demonstrate that at oblique incidence, even under controlled pump-control wave conditions, the temporal profile of $\omega_p(t)$ becomes increasingly structured as the pump intensity rises, and that a single inverted Gaussian is insufficient to capture the dynamics in strongly driven, subwavelength films. As pointed out in the main text, in reference [R20] the authors also report non-monotonic temporal variations of the plasma frequency, extracted in the linear regime from the Fresnel equations. Additional comments about similarities between our work and that of reference [R20] are found in the main text.

**Experimental Results**

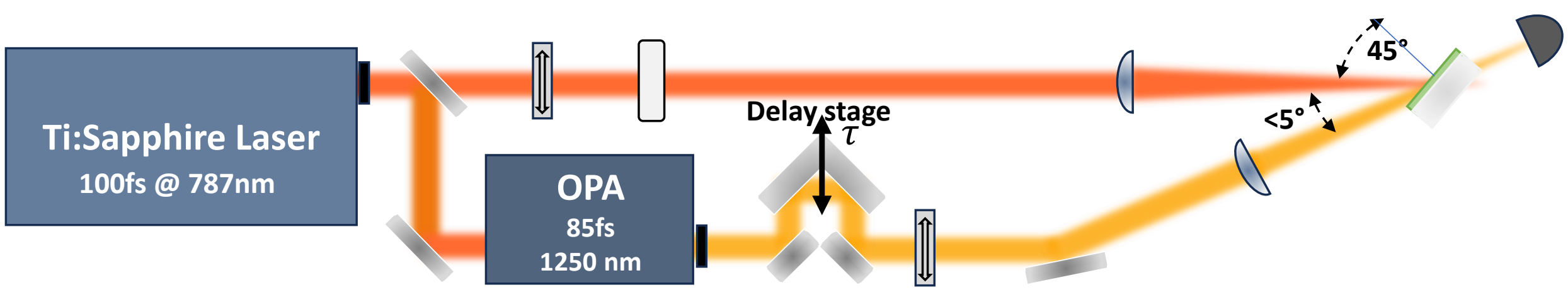

**Fig. FS2:** *Schematic of the experimental pump-control wave setup.* A Ti:sapphire laser produces ~100-fs pulses at 787 nm, used as the pump. A portion of the beam seeds an optical parametric amplifier (OPA) that generates ~85-fs control wave pulses at 1250 nm. The two beams are routed through mirrors and focusing optics, combined collinearly, and directed onto the sample under controlled angles of incidence. This configuration enables synchronized pump-control wave excitation for time-resolved nonlinear harmonic generation measurements.

The setup is shown in Fig. FS2. The measurements were performed using 10 Hz pump and control wave pulses incident on non-optimized, 240-nm-thick commercially available ITO film at angles of 45° and 50°, respectively. The pump wavelength was fixed at 787 nm and its peak intensity was 1.35 TW/cm². The control wave was tuned at 1250 nm with peak power density of 150 GW/cm². Both beams were TM-polarized, with the electric fields oriented perpendicular to the optical table. The pump and control wave spot areas at the sample were approximately 1 mm² and 0.5 mm², respectively. The second- and third-harmonic spectra generated by the control wave beam and shown in Fig.5 in the main text were collected using a spectrometer, with and without the pump present. The FWM contribution was separated from the harmonic signal spatially, as the FWM emission occurs at a different angle owing to its distinct phase-matching condition. The spectra reported here therefore correspond only to the collinear harmonic emission. All measurements were normalized to the corresponding signal from the bare substrate.

**Self-induced sideband generation in the absence of the pump**

The emergence of weak spectral sidebands in Fig.5 in the text, in the absence of the pump field, deserves a brief comment. While recent literature frequently interprets such effects within the broader framework of time-varying media, it should not come as a surprise that spectral broadening and sideband generation are natural consequences of self-consistent solutions of Maxwell's equations coupled to a time-dependent material polarization. Similar phenomena have long been recognized in nonlinear optics through mechanisms such as self-phase modulation, self-steepening, free-carrier dynamics, and plasma-induced refractive-index changes. The appearance of new frequency components therefore does not necessarily imply the existence of an externally imposed temporal modulation. Rather, it may instead arise from the intrinsic evolution of the medium driven by the optical field itself.

To examine this possibility, we simulated the propagation of the control pulse in the absence of the pump. As shown in Fig. FS3, weak but discernible sidebands emerge in both reflection and transmission as the probe intensity is increased. Their appearance is accompanied by modifications of the transient electron-plasma dynamics, indicating that the control pulse alone is capable of inducing a modest self-modulation of the material response. The effect is significantly weaker than the pump-assisted reshaping discussed in the main text, but it demonstrates that the electron plasma does not remain stationary even under single-beam excitation.

The sidebands in reflection are much stronger than those in transmission because they are mostly surface-generated and immediately ejected into the surrounding vacuum. In contrast, the transmitted spectrum has to propagate through the material while experiencing absorption, leading to the drastic reduction of its amplitude. Nevertheless, the figure demonstrates that the electronic plasma does not remain stationary even under single-beam excitation [R4]. The sidebands arise naturally from the self-consistent solution of Maxwell's equations coupled to a time-dependent polarization and therefore represent a form of self-induced spectral modulation. These results indicate that the spectral features observed experimentally in the absence of the pump can be fully explained within the same dynamical framework used throughout the manuscript, without invoking additional mechanisms. While the pump field enhances and controls the process, it is not required for the onset of the underlying plasma dynamics.

These results suggest that the weak spectral modulation observed experimentally in the no-pump configuration of Fig. 5 may originate from self-induced plasma dynamics rather than either experimental artifacts, FWM, or residual background signals. The simulations therefore support a unified interpretation in which both the pump-assisted and probe-only configurations arise from the same underlying physics: a dynamically evolving electron plasma coupled self-consistently to the electromagnetic field. The role of the pump is not to create a qualitatively new mechanism, but rather to enhance and control a dynamical process that is already present at a weaker level under sufficiently intense probe excitation.

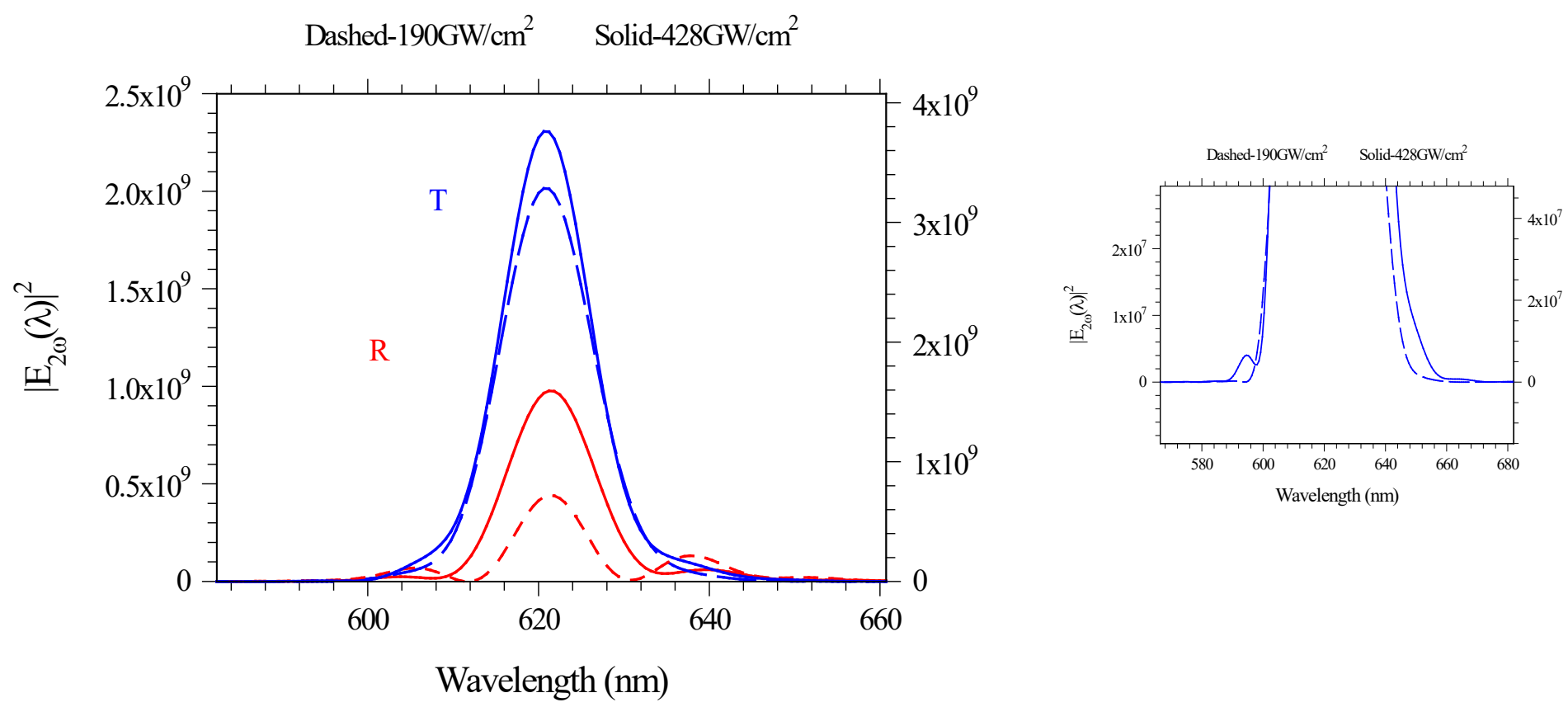

**Fig. FS3:** Left panel: Calculated reflected (R) and transmitted (T) spectra obtained using only the control pulse. Dashed and solid curves correspond to peak intensities of 190 GW/cm² and 428 GW/cm², respectively, for a given set of parameters. Clear sidebands of increasing intensity emerge in the reflected component, as weaker sidebands emerge in the transmitted spectrum (right panel). This demonstrates that the electronic plasma does not remain stationary even under single-beam excitation.

## Pump-Induced Damping Mechanisms in Time-Dependent Hydrodynamics

The hydrodynamic-Maxwell model becomes especially rich when the strong pump modifies not only the plasma frequency but also the effective mass and carrier density of the conduction electrons simultaneously [R3-R5]. These quantities enter the equations of motion through total time derivatives, and their ultrafast evolution generates additional damping-like terms that are entirely absent in static media. For simplicity, the main-text model used to produce Figs. 2 and Figs. 6 excluded these nonlinear damping contributions. Their inclusion is generally required to describe the more complex ultrafast regimes explored in Refs. [R3] and [R4], where extreme excitation conditions drive substantial changes in electronic temperature, nonparabolicity, and carrier density. However, preliminary tests and the figures indicate that the qualitative trends and physical insights of the simplified framework remain intact, perhaps because the temporal dynamics of the plasma frequency is equivalent to modulating charge density and effective mass. Because these additional terms directly influence the coherence time of the nonlinear polarization and the efficiency and timing of the emitted harmonics, they will be studied separately in this context. For now, it is useful to derive them explicitly and interpret their physical meaning within a unified hydrodynamic framework. The starting point is the free-electron momentum equation, written in a form that allows the effective mass to depend on time. In its simplest hydrodynamic form, the equation reads

$$\frac{d}{dt}\left[m_f^*(t)\mathbf{v}(\mathbf{r},t)\right] = e\mathbf{E}(\mathbf{r},t) + \mathbf{F}_{HD}(\mathbf{r},t) \qquad \text{(S8)}$$

where $\mathbf{v}(\mathbf{r},t)$ is the velocity field, $m_f^*(t)$ is the pump-dependent effective mass, and $\mathbf{F}_{HD}(\mathbf{r},t)$ collects all hydrodynamic forces, including Lorentz, pressure, convective, and nonlocal contributions. Expanding the total derivative produces two terms: one proportional to the acceleration and one proportional to the time derivative of the mass. The expansion reads

$$m_f^*(t)\frac{d\mathbf{v}}{dt} + \frac{dm_f^*(t)}{dt}\mathbf{v} = e\mathbf{E} + \mathbf{F}_{HD} \qquad \text{(S9)}$$

The second term on the left is the origin of the mass-driven damping. To see how it enters the polarization equation, recall that the free-electron polarization satisfies

$$\dot{\mathbf{P}}_f = ne\mathbf{v} \qquad \text{(S10)}$$

so that the velocity may be written as

$$\mathbf{v} = \frac{\dot{\mathbf{P}}_f}{ne}. \qquad \text{(S11)}$$

Substituting this expression into the mass-derivative term yields

$$\frac{dm_f^*(t)}{dt}\mathbf{v} = \frac{1}{ne}\frac{dm_f^*(t)}{dt}\dot{\mathbf{P}}_f. \qquad \text{(S12)}$$

When the polarization equation is rearranged into the familiar form

$$\ddot{\mathbf{P}}_f + \Gamma(t)\dot{\mathbf{P}}_f + \ldots = \text{driving terms}, \qquad \text{(S13)}$$

this contribution appears as an additive correction to the damping coefficient. Dividing the entire equation by $m_f^*(t)$ isolates the term $\Gamma_m(t) = \frac{1}{m_f^*(t)}\frac{dm_f^*(t)}{dt}$.

This expression shows that whenever the effective mass changes in time, the polarization experiences an additional friction proportional to the fractional rate of mass change. In ENZ materials such as ITO, the strong pump drives hot-electron nonparabolicity, which increases the effective mass on femtosecond timescales. The resulting term therefore acts as a genuine, pump-induced damping mechanism that shortens the coherence time of the nonlinear free-electron motion. Because high-order harmonics rely on long-lived, strongly anharmonic trajectories of the polarization, this mass-driven damping suppresses the highest orders most strongly and narrows the temporal window in which harmonic emission can occur.

A second damping mechanism emerges when the strong pump is tuned near a Lorentzian resonance and promotes electrons from the valence band into the conduction band. In this case the carrier density becomes time-dependent, in the sense that the total number of free charges increases, and the continuity equation governs its evolution. The continuity equation reads

$$\frac{\partial n}{\partial t} + \nabla \bullet (ne\mathbf{v}) = S(\mathbf{r},t)\,, \qquad \text{(S14)}$$

where S represents the pump-induced interband source term. The density enters the polarization through the relation $\dot{\mathbf{P}}_f = ne\mathbf{v}$. Differentiating this expression with respect to time produces

$$\ddot{\mathbf{P}}_f = e\left(\frac{dn}{dt}\mathbf{v} + n\frac{d\mathbf{v}}{dt}\right). \qquad \text{(S15)}$$

The first term on the right contains the time derivative of the density. Substituting the expression for the velocity in terms of the polarization gives

$$e\frac{dn}{dt}\mathbf{v} = \frac{1}{n}\frac{dn}{dt}\dot{\mathbf{P}}_f\,. \qquad \text{(S16)}$$

When this term is moved to the left-hand side of the polarization equation, it appears with a negative sign in the effective damping. The resulting contribution is $\Gamma_n(t) = -\frac{1}{n(t)}\frac{dn(t)}{dt}$. If the pump injects carriers so that the density increases in time, then $dn/dt > 0$ and $\Gamma_n(t)$ becomes negative. This term therefore acts as anti-damping, transiently reducing the effective friction on the polarization. Physically, the same current is now shared among a larger number of electrons, and the collective motion becomes easier to sustain. During the rising edge of the pump, when carrier injection is strongest, the density-driven term can partially cancel or even outweigh the mass-driven damping, producing a brief interval in which the nonlinear polarization evolves with enhanced coherence.

The total effective damping experienced by the control wave is the sum of the intrinsic Drude damping and the two dynamical contributions derived above. The combined expression is [R3-R5]:

$$\gamma_f^{eff}(t) = \gamma_f + \frac{1}{m_f^*(t)}\frac{dm_f^*(t)}{dt} - \frac{1}{n(t)}\frac{dn(t)}{dt}. \qquad \text{(S17)}$$

This formula makes clear that the pump renormalizes the coherence properties of the control wave and all its generated harmonics. If the mass grows more rapidly than the density, the effective damping increases, the nonlinear response is suppressed, and conversion efficiency is reduced. If on the other hand the density grows more rapidly than the mass, the effective damping decreases and the nonlinear response may be enhanced. Because the two processes generally occur on different timescales (carrier injection being tied to the optical cycle of the pump, and mass evolution being tied to slower hot-electron dynamics) effective damping and instantaneous conversion efficiencies become highly asymmetric in time.

This asymmetry produces a form of temporal gating. Harmonic emissions are not uniform across the duration of the control wave but is instead concentrated in those intervals where the effective damping is minimized. During the rising edge of a resonant pump, when carrier injection is strongest, the density-driven term can open a gate of enhanced harmonic emission. Later in the interaction, as carrier injection slows while the effective mass continues to grow, the mass-driven term closes the gate and suppresses further emission. The resulting harmonic bursts inherit the temporal structure of the pump-induced electronic dynamics, and their spectral signatures, broadening, asymmetry, multi-lobed features, and shifts in the center frequency, reflect the interplay between the time-dependent refractive index and the time-dependent coherence of the nonlinear polarization.

In regimes where the pump is tuned far from resonance, the density remains essentially constant and the mass-driven term dominates, reflecting and producing the purely suppressive gating described in the main text, leading to reduced conversion efficiencies. Near resonance, however, the density-driven term introduces a qualitatively new behavior [R19]: the harmonic yield may increase with pump intensity at first, reach a maximum when the density-driven and mass-driven terms balance, and then decrease again as hot-electron mass growth takes over. This non-monotonic dependence of the harmonic yield on pump intensity is a direct signature of the many competing dynamical channels that shape the nonlinear response of time-varying media that should be considered.

Together, these mechanisms illustrate that the pump does far more than modulate the plasma frequency. It reshapes the coherence environment in which the control wave generates harmonics, carving out temporal windows of either enhanced or suppressed emission and imprinting its own ultrafast dynamics onto the amplitude, phase, and spectral structure of the resulting high-order radiation. This richer picture emerges naturally from the hydrodynamic-Maxwell framework and highlights the importance of retaining the full, time-dependent structure of free- and bound-electron equations when modeling nonlinear optics in ENZ materials under extreme excitation conditions. The emergence of these dynamical damping and anti-damping coefficients highlights why standard parametric, static-permittivity models, such as instantaneous four-wave mixing or cross-phase modulation tensors, fail under extreme excitation conditions. Because parametric wave-mixing assumes a fixed electronic framework with a constant coherence lifetime, it fundamentally lacks the mathematical machinery required to capture this active, pump-governed renormalization of the medium's internal dissipative states.